# obscura: A modular C++ tool and library for the direct detection of (sub-GeV) dark matter via nuclear and electron recoils

**Timon Emken**[1]

**1** The Oskar Klein Centre, Department of Physics, Stockholm University, AlbaNova, SE-10691 Stockholm, Sweden

## Summary

The observation of a large number of gravitational anomalies on astrophysical and cosmological scales have convinced us that the majority of matter in the Universe is invisible (Bertone et al., 2005; Bertone & Tait, 2018). This *dark matter* (DM) must be fundamentally different from the known matter we can describe using the Standard Model of Particle Physics (SM). Its only established property is that it interacts gravitationally and indeed dominates the gravitational potential of galaxies and galaxy clusters. One of the leading hypothesis is that DM is made up of one or more new particles and that galaxies such as our Milky Way are embedded in gigantic haloes of these as of yet undetected particles. Our planet would at any moment be penetrated by a stream of these particles without much of an effect. If these dark particles interact with nuclei and/or electrons via some new force besides gravity, they would on occasion collide with a terrestrial particle. *Direct detection experiments* search for these kind of interactions and aim to observe DM events within a detector caused by an interaction with target nuclei (Drukier et al., 1986; Goodman & Witten, 1985; Wasserman, 1986) or electrons (Essig et al., 2012; Kopp et al., 2009). These experiments are typically placed deep underground to shield them from possible backgrounds, e.g., due to cosmic rays.

In order to interpret the outcome of direct detection experiments, we need to make predictions for the expected events caused by the incoming DM particles. In all cases, this requires making a number of assumptions about the possible particle attributes of DM (e.g., mass and interaction strength) and the properties of the galactic DM halo (e.g. the local DM density and their energy distribution) (Del Nobile, 2021; Lewin & Smith, 1996).

`obscura` is a tool to make quantitative predictions for direct DM searches, analyse experimental data, and derive, e.g., exclusion limits as seen in Figure 1. `obscura` can e.g. be used to compute the expected event rates in terrestrial detectors looking for rare interactions between the DM and nuclei or electrons. There are many different experimental techniques and targets proposed and applied for direct detection experiments (Griffin et al., 2020). Additionally, due to our ignorance about the particle physics of DM there exists a plethora of viable assumptions and models. The vast variety of viable assumptions is reflected by the modular, polymorphic structure of all modules of the `obscura` library which allows to easily extend `obscura`'s functionality to the users' new idea on the fundamental nature of DM particles, or on a new detection technology. For example, the library can handle any kind of DM particles of any mass, provided that the scattering is well-described by non-relativistic dynamics, and that the differential (nucleus and/or electron) scattering cross sections depend only on the momentum transfer, the relative speed between DM and target, and at most one additional dynamic parameter such as the center-of-mass energy or the local temperature of the target. Furthermore, a generic structure also allows applications of (a subset of) the obscura classes in a variety of DM research projects even beyond the context of direct detection, e.g., to compute DM capture rates in the Sun (Emken, 2021a).

For more details on obscura and its implementation in C++, we refer to the documentation[1].

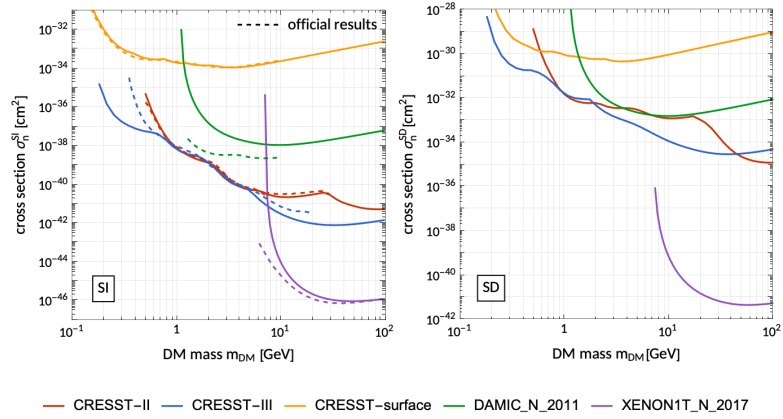

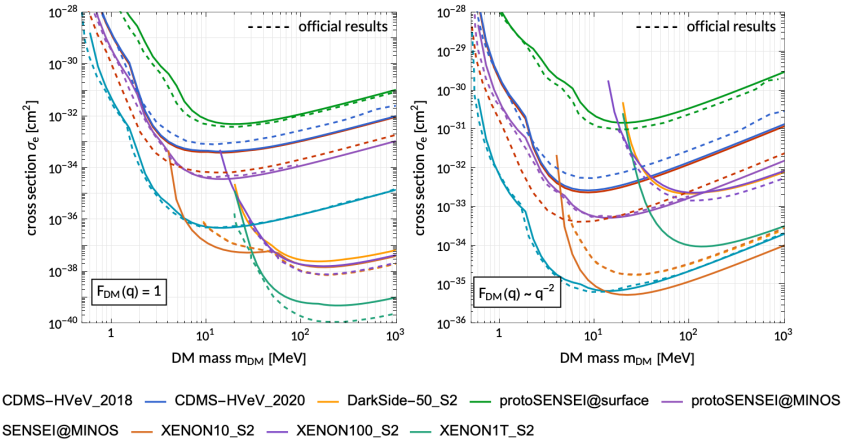

Figure 1: Excluded regions (90% confidence level) of the DM parameter space given by the $(m_{\rm DM}, \sigma_i)$ plane, where $m_{\rm DM}$ is the assumed DM mass and $\sigma_i$ is the interaction cross section with target $i$. For comparison, the dashed lines denote the official results published by the experimental collaborations. Some of the obscura results are conservative due to a simplified analysis.

# The modular structure of direct detection computations

Making predictions and performing analyses for direct detection experiments involves methods and results from statistics, astrophysics, particle physics, nuclear and atomic physics, and condensed matter physics. For each of these fields, we need to make choices and assumptions which will affect our interpretation of DM searches.

As an example, let us look at the energy spectrum of DM induced ionization events as derived by Essig et al. (2016).

$$\frac{\mathrm{d}R_{\rm ion}}{\mathrm{d}E_e} = N_T \frac{\rho_\chi}{m_{\rm DM}} \sum_{n,\ell} \int \mathrm{d}q^2 \int \mathrm{d}v \, v f_\chi(v) \frac{1}{4E_e} \frac{\mathrm{d}\sigma_e}{\mathrm{d}q^2} \left| f_{\rm ion}^{n\ell}(q, E_e) \right|^2 . \quad (1)$$

---
[1]The latest version of the documentation can be found under https://obscura.readthedocs.io.

The DM mass $m_{\rm DM}$ and the differential DM-electron scattering cross section $\frac{{\rm d}\sigma_e}{{\rm d}q^2}$ are defined by the assumed particle physics of the hypothetical DM particle the experiment is probing. The velocity distribution $f_\chi(v)$ and the local DM energy density $\rho_\chi$ are important inputs from astrophysics and cosmology. Lastly, the ionization form factor $f_{\rm ion}^{n\ell}(q,E_e)$ encapsulates the atomic physics of the electronic bound states and describes the probability of an electron with quantum numbers $(n\ell)$ to get ionized by an incoming DM particle. As we can see, the evaluation of this expression for the electron recoil spectrum is highly modular combining inputs from various fields of research. This modularity should be reflected in the structure of corresponding research software.

It is our ambition for the `obscura` code that the basic functionality does not rely on specific choices and that no particular assumption is hard-coded. Instead the basic code's setup is polymorphic and written in terms of generic base classes widely agnostic to specific assumptions. Of course, a number of standard ideas and models are implemented as derived classes, which also illustrate the usage of the base classes. The classes are described in more detail in the [documentation](#).

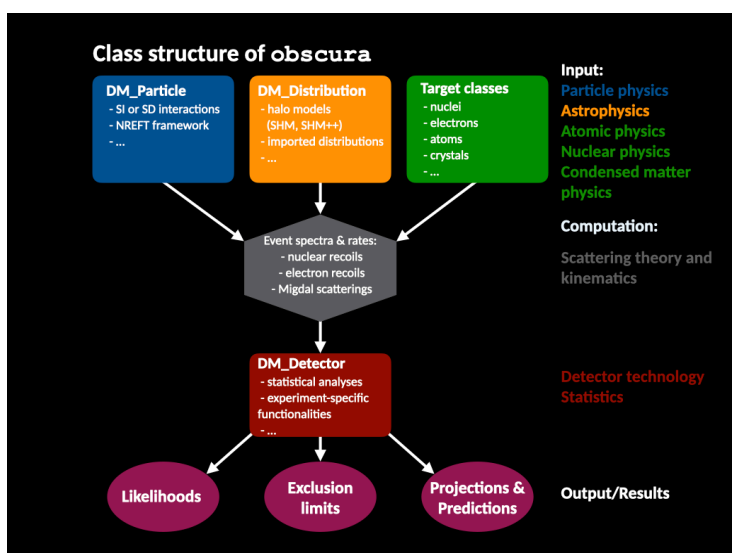

**Figure 2:** The class structure of `obscura`.

External research software can use `obscura` by implementing its classes following the dependencies indicated by the flow chart of [Figure 2](#) and computing standard quantities in the context of direct detection of dark matter. In addition, these classes are meant to be general-purpose and can be applied in other contexts depending on the research project's main objectives. It is also possible to exploit the polymorphic structure and extend its functionality by creating new derived classes based on the users' own ideas. As a final benefit of the polymorphic structure, any research software that is formulated entirely in terms of the abstract base classes can later on be used with any derived classes and allows analyses and research for a broad range of alternative assumptions without changing the core of the scientific code. As an example, the `DaMaSCUS-SUN` code uses `obscura` in the context of Monte Carlo simulations ([Emken, 2021a](#), [2021b](#)).

## Statement of need

For the interpretation of past and future direct searches for DM particles, it is important to be able to provide accurate predictions for event rates and spectra under a variety of possible and viable assumptions in a computationally efficient way. While there exists a few tools to compute DM induced nuclear recoil spectra, such as DDCalc (Athron & others, 2019; Bringmann & others, 2017) or WimPyDD (Jeong et al., 2021), obscura is not limited to nuclear targets. Instead its main focus lies on sub-GeV DM searches probing electron recoils which typically requires methods from atomic and condensed matter physics (Catena et al., 2020, 2021; Essig et al., 2016). In the context of sub-GeV DM searches, new ideas such as target materials or detection techniques are being proposed regularly, and the theoretical modelling of these are getting improved continuously (Griffin et al., 2021). At the same time, currently running experiments continue to publish their results and analyses, setting increasingly strict bounds on the DM parameter space. In such a dynamic field, obscura can be an invaluable tool due to its high level of adaptability and facilitate and accelerate the development of new, reliable research software for the preparation of a DM discovery in the hopefully near future.

# Acknowledgements

The author thanks Radovan Bast for valuable discussions and support regarding research software engineering. The author was supported by the Knut & Alice Wallenberg Foundation (PI, Jan Conrad).

# References

Athron, P., & others. (2019). Global Analyses of Higgs Portal Singlet Dark Matter Models using GAMBIT. *Eur. Phys. J. C*, *79*(1), 38. https://doi.org/10.1140/epjc/s10052-018-6513-6

Bertone, G., Hooper, D., & Silk, J. (2005). Particle Dark Matter: Evidence, Candidates and Constraints. *Phys. Rept.*, *405*, 279–390. https://doi.org/10.1016/j.physrep.2004.08.031

Bertone, G., & Tait, T., M. P. (2018). A new Era in the Search for Dark Matter. *Nature*, *562*(7725), 51–56. https://doi.org/10.1038/s41586-018-0542-z

Bringmann, T., & others. (2017). DarkBit: A GAMBIT Module for Computing Dark Matter Observables and Likelihoods. *Eur. Phys. J. C*, *77*(12), 831. https://doi.org/10.1140/epjc/s10052-017-5155-4

Catena, R., Emken, T., Matas, M., Spaldin, N. A., & Urdshals, E. (2021). Crystal responses to general Dark Matter-Electron Interactions. *Phys. Rev. Res.*, *3*(3), 033149. https://doi.org/10.1103/PhysRevResearch.3.033149

Catena, R., Emken, T., Spaldin, N. A., & Tarantino, W. (2020). Atomic Responses to General Dark Matter-Electron Interactions. *Phys. Rev. Res.*, *2*(3), 033195. https://doi.org/10.1103/PhysRevResearch.2.033195

Del Nobile, E. (2021). *Appendiciario – A Hands-on Manual on the Theory of Direct Dark Matter Detection*. http://arxiv.org/abs/2104.12785

Drukier, A. K., Freese, K., & Spergel, D. N. (1986). Detecting Cold Dark Matter Candidates. *Phys. Rev. D*, *33*, 3495–3508. https://doi.org/10.1103/PhysRevD.33.3495

Emken, T. (2021a). *Solar Reflection of Light Dark Matter with Heavy Mediators*. http://arxiv.org/abs/2102.12483

Emken, T. (2021b). *Dark Matter Simulation Code for Underground Scatterings - Sun Edition (DaMaSCUS-SUN) [Code, v0.1.0]* (Version v0.1.0) [Computer software]. Astrophysics Source Code Library record [ascl:2102.018]. The code can be found under https://github.com/temken/damascus-sun. Zenodo. https://doi.org/DOI:10.5281/zenodo.4559874

Essig, R., Fernandez-Serra, M., Mardon, J., Soto, A., Volansky, T., & Yu, T.-T. (2016). Direct Detection of Sub-GeV Dark Matter with Semiconductor Targets. *Journal for High Energy Physics*, *05*, 046. https://doi.org/10.1007/JHEP05(2016)046

Essig, R., Mardon, J., & Volansky, T. (2012). Direct Detection of Sub-GeV Dark Matter. *Phys. Rev. D*, *85*, 076007. https://doi.org/10.1103/PhysRevD.85.076007

Goodman, M. W., & Witten, E. (1985). Detectability of Certain Dark Matter Candidates. *Phys. Rev. D*, *31*, 3059. https://doi.org/10.1103/PhysRevD.31.3059

Griffin, S. M., Inzani, K., Trickle, T., Zhang, Z., & Zurek, K. M. (2020). Multichannel Direct Detection of Light Dark Matter: Target Comparison. *Phys. Rev. D*, *101*(5), 055004. https://doi.org/10.1103/PhysRevD.101.055004

Griffin, S. M., Inzani, K., Trickle, T., Zhang, Z., & Zurek, K. M. (2021). Extended Calculation of Dark Matter-Electron Scattering in Crystal Targets. *Phys. Rev. D*, *104*(9), 095015. https://doi.org/10.1103/PhysRevD.104.095015

Jeong, I., Kang, S., Scopel, S., & Tomar, G. (2021). *WimPyDD: an Object-oriented Python Code for the Calculation of WIMP Direct Detection Signals*. http://arxiv.org/abs/2106.06207

Kopp, J., Niro, V., Schwetz, T., & Zupan, J. (2009). DAMA/LIBRA and Leptonically Interacting Dark Matter. *Phys. Rev. D*, *80*, 083502. https://doi.org/10.1103/PhysRevD.80.083502

Lewin, J. D., & Smith, P. F. (1996). Review of Mathematics, Numerical Factors, and Corrections for Dark Matter Experiments based on Elastic Nuclear Recoil. *Astropart. Phys.*, *6*, 87–112. https://doi.org/10.1016/S0927-6505(96)00047-3

Wasserman, I. (1986). Possibility of Detecting Heavy Neutral Fermions in the Galaxy. *Phys. Rev. D*, *33*, 2071–2078. https://doi.org/10.1103/PhysRevD.33.2071